# How to Control a Decrease in Physics Enrollment?

*Samina S.Masood*
University of Houston Clear Lake
**Email:** masood@uhcl.edu

ABSTRACT

*Physics community is generally concerned about the decrease in Physics enrollment. Due to the deficiency of specially qualified Physics teachers, high school Physics is sometimes taught by general science teachers who may not be trained to motivate students to study Physics. These students will neither be inclined to register into Physics courses nor plan to teach Physics. They will be unable to find out the relevance of Physics with daily life, its application in different disciplines and even the job market. In this situation, we have to carefully make the existing Physics programs more attractive, instead of closing them down. We discuss teaching methodology and course requirements that will help to make Physics programs more attractive and preferable for incoming students. However, we emphasize to set our goals and plan to increase enrollment in parallel steps such as proper information about the program, help in developing a required mathematics background, offering scholarships, teaching assistantship or internships and involving them in research. Also we need to make Physics programs more interesting with the interdisciplinary courses and other electives which students may like. We will have to work on the retention rate to maintain enrollment without compromising on standards. However, we still have to develop a complete understanding of the problem and keep looking for a better solution.*

MOTIVATION

Physics is a very important subject. It plays a crucial role in technological development. Proper teaching of Physics plays a pivotal role in preparing upcoming generations for even more developed technological world of future. We need to produce qualified and trained Physics teachers who can teach new generations adequately and manage future directions of technological development. Thus the decrease in Physics enrollment is alarming and we cannot shut down Physics program due to the low enrollment and even loose the chances of getting better enrollment. Physics is an essential subject and is needed to be there to train future generations with technology and maintain our role as leaders in technological development. We need teachers who can teach new generations and we need experts who can play their role in the development of science and technology.

Due to the tremendous positive effects of Physics on the social and national character of human beings, we need to popularize it as a subject and communicate with parents and general public so that they can realize its importance and give their kids an option of learning Physics in the beginning of their school time. Physics is needed as a subject to be there because we need it to secure our future. Physics is an essential part of science and technology and all scientists and engineers have to take it. If this training starts from high school, scientist and engineer can take more advanced courses, instead of struggling with high school math. This will, not only develop better scientists and engineers but will encourage the interdisciplinary research. Therefore, even if we do not worry about producing a large number of physicists, we still need a large number of qualified Physics teachers who could motivate students into science and engineering disciplines and train them for future challenges. Closing Physics programs is not going to take us anywhere. Because of this strategy, Physics is only serving as a service course for other science and engineering disciplines. Enrollment in Physics program is generally going down. Instead of working on increasing enrollment, administration and higher authorities in academia are closing down Physics program, labelling it as a low producing program. Existing Physics programs are running on international students, with a large majority of American Asians. Most of these students will either go back to their countries and/or they will have to be hired in US, but they will still have difficulty getting security clearance for sensitive jobs. We are not sure if we really want to give away the sensitive jobs to other nations. Even if they want to be teachers, they will have to go through a tedious hiring procedure. In that case they may not be absorbed into the US, because if we close the programs, we will be unable to hire enough teachers and it will be a cyclic effect. We will not have enough qualified teachers who can adequately teach Physics and motivate students to study Physics. It will become a bigger challenge with the increasing number of high schools and with the growth in population.

In this paper, we give a qualitative analysis based on interaction with students from culturally diverse backgrounds and highlight those social and cultural values that can deeply affect the young population and their priorities in life. Also the educational strategies and teaching methods have to be modified according to the needs of enrolled students, their goals and the demographic mix of the class. It is a very complicated problem and has to be resolved step-by-step and keep watching its impact.

## IDENTIFICATION OF THE PROBLEM

In this era of fast paced life, well-accepted subjects that are very involved such as mathematics and Physics are losing their popularity because it needs more commitment of time, focused study, regularity, rigorous training and a real aptitude. Above all, these difficult programs do not lead to high-enough paying jobs. People may earn much more money and find more job opportunities by investing less time and work. So a majority of students are taking introductory Physics courses as required courses to complete non-Physics degree requirement only. Physics program is losing popularity and is accepted as an involved subject but does not have an adequate job market. So Physics programs are shutting down because of the low enrollment and merging with popular departments and only offer service courses.

The decrease in the Physics enrollment is the biggest challenge for physicists of this century and a number of parallel techniques are being developed to increase enrollment. There is considerable success in increasing enrollment for Physics teachers through federal projects for STEM (Science, Technology, Engineering and Mathematics) teachers such as 100K-10, PhysTEC, Uteach, etc. It also proves that we have enough students who have the capability of learning Physics and even will succeed in a Physics career, but there is a lack of motivation. A well-planned plan of study, appropriate training, job opportunities, networking between students and faculty and also meeting active researchers and well-known physicists give them role-models in the field and convince them to accept the challenges related with any field of study. Young students have a lot of energy and spirit to work hard to achieve their career goals as long as they have strong motivation.

It is also an accepted reality that the current generations with the current social structure have a weak support system and face big challenges in time management, which affects their learning ability. A large percentage of students do not choose to take time consuming subjects as they cannot find enough time to get involved in detailed study and proper understanding of any subject. They have to develop a quick learning strategy to pass their courses and complete the degree requirements. So they may have to avoid taking more involved courses, even if they like the subject. This issue is leading towards much bigger problems for future generations. With these trends of education, we have reached the point where it may become difficult to maintain the same rate of development. We can probably compromise with the temporary phase of low enrollment and work on improving standards as well as helping students resolve their issues, instead of closing a program, due to low enrollment. We may have to compromise on some reduction in profit, temporarily providing funds to low producing programs to support students and then regenerate those funds through increased enrollment. It will be a profitable investment, not a financial loss for educational institutes.

Different nations, because of the social and cultural basis of every nation, their social network, family structure, and financial aid and scholarships programs and how their youth approaches towards life, take these challenges differently. A few countries are actually doing very well with the promotion of various subjects by offering scholarships to students to send them to US and other developed countries for higher studies. A reasonable ratio of these international students chose to learn Physics and will help with the survival of Physics programs in the US. Most of the domestic US students work part-time or even full-time. They do not have any financial sharing or family support system and have to worry about paying their bills and supporting their families during their college education. They cannot afford to get into a more involved and challenging degree program, even if it pays it off later. They have a passion for these subjects. So the image of Physics being a difficult subject that requires very hard work to understand the subject and needs to learn a lot of mathematics eclipses its attraction due to personal situations.

Slimming down of enrollment is a big issue and educational institutes are genuinely concerned about these low enrollment programs because they are not as productive as other high enrollment programs are. Physics does not even give the same kind of salaries. However, we have to plan our future properly. We then have to choose between our long term and short term goals. For this purpose we need to look at some aspects of this problem in order to get to the root-cause of the issue. We discuss the problems one-by-one in detail.

## I. Fear of Program Closing

Physics is usually a low enrollment program and its introductory courses are extensively used as service courses for other science and engineering disciplines. Introductory Physics courses are taken by all science and engineering majors. However, we usually do not have many Physics majors and departments that can meet their financial requirements while being profitable programs for campuses. This program is not an immediately paying program; however, it is an excellent place for future investment. Providing more resources to prepare good teachers and scientists is equivalent to invest in building a technologically superior nation for the future while resolving the issue of low enrollment. However, before making any decision to close down these programs, we need to look at certain factors. If we keep shutting down a highly needed program due to the low enrollment, there is no way that we can succeed at increasing enrollment.

Fear of program closure is taking a lot of faculty time, repurposed towards student recruitment. Sometimes they have to compromise on standards and favor enrollment over standards to save the program and their jobs.

We need Physics teachers at all levels. Creating good teachers is like building a good nation. These teachers can be trained in the Physics programs. If the Physics departments are squeezed down and we keep closing down these programs, how can develop special courses for Physics teachers to fulfil our future needs?

We cannot have and never have had large enrollment in Physics. We do not even need too many Physicists. We need good Physicists who can (directly or indirectly) play a key role in the development of future technologies.

As long as the Physics program can pay for its survival through service courses, there is no point in closing down the Physics program. Producing smaller number of Physics majors is still better than not producing any. We can work on getting more students in the program and not saving the program.

*Keeping standards of education and even improving the quality will provide leading physicists and great engineers. Closing down the program will not resolve any issue.*

## II. Politics in Academia

Politics is a well-known part of academic culture. People need money for research and getting money to fulfill your research needs still involves money and politics. In the distribution of small amount of funds to big divisions, Physics usually suffers for being a small department. Salary structure in academia is pretty non-uniform. It totally depends on how good you are in negotiation at the time of hiring or how good you relate with the administrators.

Individual efforts in developing a program or attracting students to the programs are usually disregarded by the administration. Research is not appreciated or encouraged, especially in teaching campuses.

Job security and teaching merits are only evaluated on the basis of instructor evaluations whereas it has been noticed that usually good researchers and busy scientists do not get very high evaluations. The evaluation procedure is designed for popularity of courses and does not take into account the instructor's role in raising the standard of education or getting more students in the program. So the teaching impact on the program is ignored compared to students liking and disliking the course. Physics instructors cannot give an easy A grade or make the course understandable for everybody without compromising on standards. Situation is much worse in public schools. Private schools have different policies in evaluations and have better standards.

Efforts of getting good evaluations may not leave the room to attract enthusiastic promising Physics students to the program.

Additionally, if faculty members try to satisfy enthusiastic students by answering their questions and encouraging them to pursue Physics as a subject, they have to pay a price and students may think that they are probably not disorganized in lectures. Instead of having well-prepared power-point presentations or lecture notes, going through mathematical steps on the board with the involvement of students actually helps students learn the mathematical techniques as well as the physical concepts

In the effort of attracting more students and helping them to be able to learn Physics themselves, it is important to make them familiar with the textbook. Some teachers put in personal efforts to use book in the class. Students may take teacher's efforts as unpreparedness. All of these individual efforts of faculty can hurt their evaluations as regular students do not like it and it will be used against them when it comes to merit selection and promotion because administrators can highlight that part of evaluations by un-interested students which serve their purpose and ignore all the achievements.

If there is any credit of program development, it tends to go to the chairs exclusively, which is also a major discouragement for regular faculty. Upper administration does not recognize individual efforts. This is a big difference between US system and other countries and is potentially affecting the standards as well.

Students handling and resolving their enrollment issues are also sabotaged by office politics and secretarial mishandlings.

Therefore, a lack of appreciation of research, ignorance of the successful efforts in bringing students in to the program, distribution of internal funding between disciplines and individuals more politically than their genuine needs and expected output and giving un-necessary favor to the course assignment are some of the issues which bring in all of the politics in academia and it becomes an integral part of academic life. It reach the level that regardless of the lack of knowledge of the subject or state of the art of the subject, administrators trained in outside discipline, interfere in the teaching strategies or teaching/research labs, which discourages students. That is where faculty needs independence in teaching and partial administrative powers to handle individual students, to satisfy their needs and attract them to the program. Administrators can play with the loop holes in the system to make use of the situations. This is where politics may have adverse effects on programs and sometimes little things become un-necessarily complicated and cost money and time to students. Sometimes personal liking and dis-likings take over the workplace interest. It has been noticed that some of the obvious human issues including

- Gender discrimination
- Racial discrimination
- Color discrimination
- Age discrimination
- Qualifications discrimination
- Ethnic background discrimination
- Sharing of resources
- Fear of higher qualification of juniors, etc.

These general approaches are integrated with the hiring process or even promotion process and/or salary determination. However, they can all be covered under the umbrella of departmental requirement and sometimes instructor evaluations can be used for covering purposes. All their achievements and contributions can just be hidden under the cover by highlighting one bad evaluation. On the other hand it has also been seen that just one contribution is highlighted so positively that an ordinary faculty member can become a star right away.

### III.    Students' Recruitment/Advising

Students' recruitment and advising strategies can change the structure of a department. Appropriate recruitment strategies can not only increase enrollment but also improve standards of the department. Recruitment goes through several stages. Admissions office is the first place where students usually start the process. Admission counselors have to be fully knowledgeable about every program. The Physics program starts facing mishaps at that initial stage. It can be commonly expected that the Physics program is not properly introduced at that level. Admission counselors may not be very specific about the program and portray Physics as Mathematics. Students may then be repelled by too much mathematics in Physics. Also the job market and salary range issues make it attractive compared to some of the other subjects. Having knowledgeable admission advisors helps in increasing enrollment.

Faculty is involved in students' recruitment differently. Undergrad recruitment is done from

- High schools
- Community colleges
- General community

Different lectures and events can be arranged to introduce Physics program, its uniqueness, benefit of learning Physics and its need for future societies. In this process sharing success stories by alumni makes everything more interesting. Again, individual efforts of faculty are usually not adequately appreciated, unless they are backed by the department chair. It means that if faculty find an opportunity to recruit students from scientific meetings or some other educational gatherings where they participate individually, they are not credited for it at all.

Faculty plays a major role in retention rates through advising. Department faculty has to figure out if a student has enough technical background including students' math and Physics introductory courses should be passed with a decent grade. If required courses are not done, what will they need at the minimum to get started with the program related courses? Too many foundation courses can be frustrating for students. Therefore faculty has to be at the helping end and help out students to develop quick and adequate foundation. This is a real challenging situation to satisfy a student without compromising on standards and basic requirements. It is just like a doctor who may have to convince young patients to get treated by bitter pills and help them accept it because of their encouraging attitude and sympathetic behavior. Faculty has to convince students that they are important and faculty care about students more than anything else. It has been repeatedly noticed that faculty face opposition from office staff, even in this situation as it may increase their work. They don't realize that student enrollment and their retention makes their jobs secure. Pleasant behavior during the time of difficult decisions is what makes students feel at home. At this point communication skills and considerate behavior with a sympathetic attitude makes students confident. However, office politics and lack of communications in between offices usually creates un-necessary issues and the student community can easily damage the reputation of the department and cause wastage of time for the office and faculty.

## IV. Research in Physics

Research is a big part of students training programs. Young minds are attracted to exploration and critical thinking and they happily accept challenges to enjoy the excitement of success in finding something new. It helps in developing an interest in the subject. As an initial step to involve students in research we may ask them to write a term paper on one of the relevant topics of the course and give class presentations. In the process of the research on a given topic they can

- Explore their own area of interest in the subject and pick up their topic of interest
- Develop the ability to focus on the topic of interest, get information and organize it
- Write it as a research article in their language.
- Present their information to the class in a convincing way and answer questions.
- Develop confidence about their knowledge and learn the skill to share information with others in a convincing way

Encouragement of research in Physics is to prepare for the future as well as to bring in research money for the survival of the program. It is a cycle: Well-trained teachers will succeed in developing interest in the subject for upcoming students and good researchers will produce good scientists and engineers at the top of grant money. Good researchers are the ambassadors of campuses and their presence has the ability to attract students to programs. Students feel proud of their teachers' research and publications. They feel honored to be taught by professors of scientific repute.

Involving students in research is another way of increasing enrollment. A very small amount of research assistantship, conference travel, paper presentations and abstract publications are some of the effective ways of student encouragement and increases enrollment. This strategy has been regularly used for eight years. Out of a small group of 10-15 students 1 to 2 are new students in the Physics program. Usually they either change major or add Physics as a dual major. It means the success rate is 10-20 percent from the introductory courses of Physics. And all of this is done successfully by using a total amount of around $5000 altogether in eight years.

Along with all of these special issues with Physics, incoming students are always pushed away from Physics due to the general impression of the subject.

## V. General Perception

In this situation, when Physics programs are being chopped up by double edged swords. Shutting down of a program takes away the opportunity to fix the enrollment issue as well as it puts another stamp on the loss of popularity of Physics. Young students can only choose to join a program if at least one or more of the following conditions are satisfied:

- Large enrollment indicating the popularity of a program
- A better job market which can be related to success stories of alumni of the program
- Easy degrees which can be earned with minimum effort and hard work while still paying.
- Large faculty size indicating the need of Physics teachers and researchers
- Faculty research and recognition in the scientific community
- Availability of scholarships or other fellowships or assistantships
- Demographics of the campus including job opportunities and internships in the region

# INTRODUCTORY COURSES IN PHYSICS

Decreasing enrollment is a challenge for most of the Physics programs and sometimes Physics faculty is compelled to compromise on standards for several reasons. Moreover people have started working on finding some alternative arrangements which may help to increase enrollment. Several Physics courses are offered in different levels to freshmen and sophomore.

- Conceptual Physics
- General Physics
- Physics for Non-science Majors
- Introductory Physics for Physics teachers
- Physics for Biology majors
- Algebra Based Physics (College Physics)
- Calculus Based Physics (University Physics)
- Physics for Scientists and Engineers

Depending on the curriculum requirement of the major and the student's background, STEM majors and STEM education majors (teachers) have to take at least one of the above sequence. Students may enroll in any of these courses at the freshmen level. However, most of them will wait until later to enroll in Physics courses to manage their grades and for math preparation. It is actually an appropriate decision if they choose to prepare for the course with sufficient background.

In the above list, the first three sequences are for non-science majors. These courses are very important for recruitment of students in Physics and engineering program. If these courses are made interesting enough, students get another opportunity (after high school) to discover their aptitude and interest in science and engineering disciplines. *However, when students take Physics at junior level, they usually do not have enough time to think about the change of major, even if they want to.*

On the other hand, science majors need specialized Physics courses such as Physics for Biology majors and even Physics for Chemistry majors. If the number of students is not very large, both of these courses can be merged into one course. This special course should be designed as the application of Physics to Chemistry and Biology. This course should include but not limited to

**Biophysics**

- Mechanics of body movement
- Work-energy relationship in biological systems
- Electromagnetism of nervous system
- Fluid mechanics of fluid motion in the living systems
- Electrolytic motion inside the cells

- Biochemistry of cellular growth and nutrition
- Chemical Physics of bonding and study of complex molecules such as proteins and lipids

Associated labs with this course have to be designed separately and include some experiments which can give an idea of physiology and anatomy also.

Class demonstrations and simple class projects (Theoretical and experimental) will make the course more interesting

Advanced topics of Physics such as Electrodynamics and quantum mechanics can be mentioned briefly to increase the interest of students.

However, it is known that the young minds are more curious and anything which can help to encourage curiosity and answer some of the questions of the inter-disciplinary topics is a very effective method to attract students. It has to be noticed that involved disciplines, such as Physics can only be manageable for students who are willing to work hard and commit for the subject. Therefore, Physics enrollment cannot be increased by just attracting students. If we enroll students who do not have enough technical background or do not have enough urge to learn an involved subject, they will not stick to the discipline. They may start it to hit a better job market but will change as soon as an opportunity comes. So we should probably look for those students who develop interest in the subject and then want to adopt it as a career.

College Physics and University Physics courses are traditionally offered to mainly freshmen science majors based on their mathematical background and degree requirements. Among these courses, College Physics is an excellent course to recruit Physics majors. These students have a science background and can browse through the Physics course and figure out their own real motivation during this exploration process. This is the only place where all of the misconceptions about Physics as a subject could be resolved and students can be recruited from undecided majors (inclined to science and engineering). Chemistry majors can actually be dual majors from other disciplines. Sometime students in College Physics may be able to take University Physics or students in University Physics may just need College Physics. However, based on the availability of the courses, who is teaching them and the requirement of the degree plan, students may register into any one of them. However, both of these course sequences can attract students in the Physics major.

Physics for Scientists and Engineers is the most specialized introductory Physics sequence. Usually, if science and engineering students have already taken Advanced Placement Calculus in High School and have some Physics background such as Advanced Placement Physics courses from high schools, they may go straight to the Physics for Scientists and Engineers. They can improve their math skill with Advanced Calculus courses in the same semester. It may be a little challenging, but still do-able. Otherwise, typically students take calculus before they start introductory Physics courses.

# NEW METHODS OF TEACHING PHYSICS

All of the physicists are aware of these problems but are unable to resolve them. Lack of administrative support aggravates these issues. Regardless of all challenges, physicists are not giving up. New innovative ideas are proposed and proper implementation of these ideas is desired. However, if we are not conscious, we may plan to increase enrollment at the cost of some long term damages. Therefore, we have to carefully define our short-term plans and long-term goals. New teaching methods are one of those challenging steps which can be extremely useful, if implemented properly. However, new courses are designed to fulfill the individual needs of a department so that they do not hurt the departmental structure. In this situation, we need to remember that Physics majors have to be trained technically with strong mathematical background so that they can develop the ability to understand Physics through mathematics and learn to interpret mathematical equations and conceptually visualize the Physics behind it. Mathematics is a language and computer provides a tool box for physicists. Higher level mathematical skill is still a requirement and we cannot compromise on the technical skills. Different courses and different teaching styles attract students of relevant aptitude and may lead to increase in enrollment. However, we need to keep in mind that our little tools will only work if they are used in a correct way, just as we cannot loosen a screw with a can-opener or cannot open a can with a screw-driver.

All of the above mentioned courses are important and different teaching styles such as studio type courses, inquiry based method, traditional lectures and labs and problem solving strategies are all different adoptable methods. None of them is either perfect or obsolete. Different students like different styles and feel at home with different methods. Work engagement and scheduling issues may also make students choose one method over the others.

Traditional Physics teaching style is to use chalk-board in the classroom and teach Physics using mathematical equations and solving them in the class step-by-step. Students are trained with the problem-solving skills in the classroom. Learning how to interpret equation and deduce physical results from mathematical expressions is a part of training. Traditional way of teaching calculus based Physics and Physics for scientists and engineers prepare students for intermediate level undergraduate Physics courses. Video analysis in the experiments and class demonstration are still good techniques to make these courses more attractive to Physics students.

Teaching Physics in Studio type setting is an approach to teach students conceptually as well as giving them a flavor of mathematics. This approach is an excellent approach for algebra based Physics students. These students are not fully prepared for learning Physics through mathematical equations. Studio type setup gives them the physical concepts and experimental testing simultaneously which makes these courses interesting enough to attract more students in Physics. In the studio type setup, critical thinking is developed and students can explore about their interest in Physics. Freshmen with undeclared science major in these courses will pick up potential Physics students in these courses. These courses are usually taken by biology, geology, environmental science and information technology majors also. Therefore the conceptual study is very important and mathematics techniques are not essential. We may recruit student from this group who may choose Physics as a second major if they find this field interesting. Inquiry based techniques are generally very effective.

Modeling in Physics is an excellent method for teacher's training. Physics teachers learn physical concepts and take new innovative ideas to their classrooms. These modeling methods are even used in professional development workshops. Conceptual Physics course and Physics for Teachers are the main courses which can effectively be taught using these modeling techniques.

Specially designed courses for non-physics majors are very interesting courses in Physics and will be very helpful to develop inter-disciplinary degrees. Again, if they are offered appropriately, they provide an excellent training for inter-disciplinary study. Courses such as biophysics, radiation-physics, chemical-physics and other applied physics courses are good trends to promote inter-disciplinary research.

Sometimes, in large departments, multiple sections of algebra based Physics courses are offered. If the majority of the enrolled students are from a specific major, Physics instructors include more examples and problems on a particular topic. In that situation, these types of college Physics level courses are named as Physics for biology majors or Physics for environmental science majors. In such courses, standard of the courses have to be maintained. However, examples and applications even the experiments involve the corresponding subjects. However, the inter-disciplinary courses are to be very carefully designed so that they do not compromise on standards. Developing such courses will promote inter-disciplinary research and encourage students to innovative approach.

## CONCLUSIONS AND RECOMMENDATIONS

Now we have several questions to answer such as what is the best way to attract students to Physics. Will teaching Physics in a different way will solve the problem or not? What are we getting out of it and what price we are paying? What else we can do to increase enrollment? Answers to these questions are not simple as they are entangled issues in between themselves. Latest studies show that there is a success in the increase in enrollment, based on teacher training programs, which is definitely encouraging and we see some light for the students' enrollment issue in Physics. However, there are a few important points under consideration.

- The new introductory courses in Physics are very important and interesting and attracting students to Physics courses. But these trends may not be equally effective to train good future physicists in all fields of studies.
- We have to clearly define our goals. Do we want to simply increase the enrollment in the introductory Physics courses or we want to produce competent physicist in next

- generations? We can probably get more Physics majors from large introductory courses, knowing what we are really looking far.
- Finding an appropriate mathematics background for Physics major is one of the big challenges for recruiters as well as for the interested students.
- Instructor evaluation procedure includes certain points, which does not evaluate a quality teacher, they evaluate a dutiful teacher. It is a discouragement for those teachers who want to work for the development of the program but at the same time, want to keep their jobs. This issue has to be handled, very carefully.
- Private campuses keep more quality researchers then teachers and their evaluation procedure is different. Therefore they do not have the same problem. However, a large percentage of students cannot afford to go to private campuses anyway. Teachers evaluations and the pressure of having a good passing rate ties teachers hands to take any positive steps to increase standards of education and or attract more students in Physics. They have to follow the guidelines to pass more students in their classes.  High school teachers are more coordinators than teachers. Evaluation procedure has tied up their hands due to keeping up with the passing rate
- Knowledge is not so much valuable. Education is business and teachers have no free hand to promote learning procedures and motivate them to go to more involved subjects at K-12 level. And it is usually too late for college teachers to do much for that.
- In some of the ethnic communities, parents decide a career for students based on the job market and salary ranges. This takes away the freedom of choosing their field of study from a large group of promising Physics students who come from such communities in US.
- With the strict classroom rules, students are not encouraged to ask questions. Too much obedience reduces free hand mental growth and damages creativity and innovative approach.
- Lack of mathematics background and late enrollment reduces the chances of getting Physics majors in colleges. So this issue has to be dealt at K-12 level effectively.
- Awareness of general public about the importance of Physics will be very helpful to resolve the enrollment issue as well as create more jobs and better financial support for students.
- In addition to the federal and state grants, public grants are needed for Physics and different foundations should work with Physics departments to establish more grants.
- Big universities and Ph.D. granting institutes are usually given priorities for STEM grants whereas small campuses need it more. This issue is more damaging for the research grants as small teaching campuses do not get research grants and it discourages research there. Even if the faculty is involved in research, they cannot involve students without funding.
- Travel grant for students to go to conferences is another good attraction for students in the program.

**Further Reading: Some of the other works of the author.**

1. Samina Masood,` How to get more students in Physics' BAPS.2013.TSF.D1.2
2. Samina Masood, `K-12 Teaching and Physics Enrollment', arXiv:1403.5501.
3. Samina Masood,` How to get more students in Physics' BAPS.2013.TSF.D1.2
4. Samina Masood, 'Why Physics Programs are Slimming Down', BAPS,2012. March, K1.007.
5. Samina Masood, `Physics Program at Univ. of Houston Clear Lake', (Presented at APS-PhysTech Meeting in Austin,(May 2011).
6. Samina Masood, Physics education at Univ.of Houston Clear Lake, `APS March Meeting in Dallas ( March 2011)
7. Samina Masood,` Improvement in High School Physics Teaching will increase Physics Enrollment' (15 April 2008 in APS April Meeting in St. Louis, Missouri)
8. Samina Masood, K-12 Teaching and Physics Enrollment, Bull.Am.Phys.Soc.52: R1.265, (2007).
9. Samina Masood, `Better Physics Teaching Can Increase Physics Enrollment', arXiv: hep-Physics/0702089
10. Samina Masood, `The Decrease in Physics Enrollment`, arXiv: Physics/0509206.